\RequirePackage{fix-cm}
\documentclass[smallextended]{svjour3}       
\smartqed  
\usepackage{graphicx}
\usepackage{enumitem}
\setlist[itemize]{label=\textbullet}

\PassOptionsToPackage{hyphens}{url}\usepackage{hyperref}
\usepackage{breakcites}
\usepackage{natbib}

\usepackage{bibspacing}
\begin{document}

\title{Data Science vs. Statistics: Two Cultures?
}


\author{Iain Carmichael       \and
        J.S. Marron 
}


\institute{I. Carmichael  \at
	      B30 Hanes Hall \at
              University of North Carolina at Chapel Hill\\
              \email{iain@unc.edu}           
           \and
           J.S. Marron \at
           352 Hanes Hall \at
           University of North Carolina at Chapel Hill\\
           \email{marron@unc.edu}
}

\date{Received: 4 January 2018 / Accepted: 21 April 2018}

\maketitle

\begin{abstract}
Data science is the business of learning from data, which is traditionally the business of statistics. Data science, however, is often understood as a broader, task-driven and computationally-oriented version of statistics.  Both the term data science and the broader idea it conveys have origins in statistics and are a reaction to a narrower view of data analysis. Expanding upon the views of a number of statisticians, this paper encourages a big-tent view of data analysis. We examine how evolving approaches to modern data analysis relate to the existing discipline of statistics (e.g. exploratory analysis, machine learning, reproducibility, computation, communication and the role of theory). Finally, we discuss what these trends mean for the future of statistics by highlighting promising directions for communication, education and research.

\end{abstract}

\section{Introduction}
\label{intro}

A simple definition of a data scientist is someone who uses data to solve problems. In the past few years this term has caused a lot of buzz\footnote{\url{https://hbr.org/2012/10/data-scientist-the-sexiest-job-of-the-21st-century}} in industry, questions in science\footnote{\url{https://www.wired.com/2008/06/pb-theory/}} and consternation in statistics\footnote{\url{https://simplystatistics.org/2015/10/29/the-statistics-identity-crisis-am-i-really-a-data-scientist/}}. One might argue that \textit{data science} is simply a rebranding of \textit{statistics} (e.g. ``data science is statistics on a Mac"\footnote{\url{https://twitter.com/cdixon/status/428914681911070720}}), but this comment misses the point. 

While data analysis has been around for a long time, its economics (costs of and value derived) have changed primarily due to technology driving the availability of data, computational capabilities and ease of communication. The most obvious advances are in computer hardware (e.g. faster CPUs, smaller microchips, GPUs, distributed computing). Similarly, algorithmic advances play a big role in making computation faster and cheaper (e.g. optimization and computational linear algebra). There are many new/improving technologies which allow us to gather data in new, faster and cheaper ways, including: drones, medical imaging, sensors (e.g. Lidar), better robotics, Amazon's Mechanical Turk, wearables, etc. Finally, improved software (e.g. see Section \ref{sss:LP}) makes it faster, cheaper and easier to communicate the results of data analysis, distribute software, and publish research/educational resources. 

These changes mean that more people get more value out of data analysis. One would be hard pressed to find an industry or academic discipline not impacted by data analysis. These developments also mean that data analysis has become increasingly multidisciplinary and collaborative. The field of data analysis has broadened in that more people want to analyze data and data analysis draws on more disciplines. Areas of statistics previously considered specialized (e.g. statistical software, exploratory analysis, data visualization, high dimensional analysis, complex data objects and the use of optimization methods) have become dramatically more valuable.

Various viewpoints have been expressed on the current state/future of statistics and the fashionable topic of \textit{data science} \citep{tukey1962future, wu1998statistics, cleveland2001datascience, breiman2001statistical, hand2006classifier, yu2014let, wasserman2014rise, efron2016computer, donoho201750, hooker2017machine, buhlmann2016mathematics, blei2017science, barocas2017social, buhlmann2018statistics, reid2018statistical}. Views have also been been expressed in blogs: Simply Statistics\footnote{\url{https://simplystatistics.org/}}, Statistical Modeling, Causal Inference, and Social Science\footnote{\url{http://andrewgelman.com/}}, and Normal Deviate\footnote{\url{https://normaldeviate.wordpress.com/}}. 

This paper offers additional opinions and perspectives and we strive for a unique viewpoint through combining both new and old ways of thinking. The first author is a computationally oriented PhD student in a statistics, who has developed and taught an entirely new undergraduate course on data science\footnote{\url{https://idc9.github.io/stor390/}} for his department. The second author has 35 years of academic experience in statistics including: research, collaborative applications and teaching at all levels. He has offered previous opinions on \textit{big data}, in particular on the topic of \emph{robustness against heterogeneity} in \cite{marron2017big}.

In Section \ref{s:datasci} we discuss the definition of \textit{data science} and how it relates to the discipline of \textit{statistics}.   In Section \ref{s:modes} we discuss a few \textit{principal components} of data analysis which give insights into aspects of modern data analysis (e.g. why the rise of computation is tied to the rise of exploratory and predictive analysis). Finally, in Section \ref{s:forward} we discuss future directions of statistics research, communication and education including: complex data, robustness, machine learning and data processing, and literate programming.

\section{What is data science?}
\label{s:datasci}

To a general audience, data science is often defined as the intersection\footnote{This idea is usually communicated through a venn diagram e.g. \url{http://drewconway.com/zia/2013/3/26/the-data-science-venn-diagram}.} of three areas: math/statistics, computation and a particular domain (e.g. biology) \citep{conway2010data, yu2014let, blei2017science}. Implicit in this definition is the focus on solving specific problems (in contrast with the type of deep understanding that is typical in academic statistics\footnote{\url{https://simplystatistics.org/2014/07/25/academic-statisticians-there-is-no-shame-in-developing-statistical-solutions-that-solve-just-one-problem/}}). The focus on problem solving is important because it explains differing judgements to be found on how to value contributions to the field. As stated in \cite{cleveland2001datascience}

\begin{quotation}
[results in] data science should be judged by the extent to which they enable the analyst to learn from data... Tools that are used by the data analyst are of direct benefit. Theories that serve as a basis for developing tools are of indirect benefit. 
\end{quotation}

For the purpose of this article we define data science as the union of six areas of \textit{greater data science} which are borrowed from David Donoho's article titled ``50 Years of Data Science" \citep{donoho201750}. We will refer to these as GDS 1-6 and more details can be found in Section 8 from Donoho's article\footnote{\url{http://www.tandfonline.com/doi/pdf/10.1080/10618600.2017.1384734}}.

\begin{enumerate}
\item Data gathering, preparation, and exploration
\item Data representation and transformation\footnote{This includes both databases and mathematical representations of data.}
\item Computing with data 
\item Data modeling 
\item Data visualization and presentation 
\item Science about data science\footnote{For example, reproducible research would fall under this category and point 5.}
\end{enumerate}

The purpose of defining data science in this way is to (A) better capture where people who work with data spend their time/effort and (B) put more focus on the value of each tool for providing insights. This definition is given in contrast to the current, perceived state of statistics. A number of other statisticians have proposed similar definitions using different terminologies and ontologies \citep{tukey1962future, chambers1993greater, cleveland2001datascience, yu2014let}. People might reasonably tweak the above definition.

The term \textit{data science}, both the literal string and the broader idea that it conveys, originates from statisticians (see Section \ref{ss:name}). In a 1993 essay titled ``Greater or Lesser Statistics: a Choice for Future Research" statistician John Chambers wrote \citep{chambers1993greater}

\begin{quotation}
Greater statistics can be defined simply, if loosely, as everything related to learning from data, from the first planning or collection to the last presentation or report. Lesser statistics is the body of specifically statistical methodology that has evolved within the profession - roughly, statistics as defined by texts, journals, and doctoral dissertations. Greater statistics tend to be inclusive, eclectic with respect to methodology, closely associated with other disciplines, and practiced by many outside of academia and often outside professional statistics. Lesser statistics tends to be exclusive, oriented to mathematical techniques, less frequently collaborative with other disciplines, and primarily practiced by members of university departments of statistics.
\end{quotation}
We take the position that data science is a reaction to the narrow understanding of \textit{lesser statistics}; simply put, data science has come to mean a broader view of statistics.

It's worth noting that our discussion is about data science from the perspective of statistics. Since data science is a multidisciplinary field, other disciplines, such as computer science, might see data science in a different way. For example, computer science might focus more on: machine learning, large scale computation and data storage/retrieval.

\subsection{What's in a name?}
\label{ss:name}

The term data science has caused excitement, confusion and controversy. Some of the confusion is from the lack a of a consistent definition\footnote{We take the position that data science is the practice of broader statistics.}. There is an ecosystem of related terms (e.g. \textit{analytics}, \textit{business intelligence}, \textit{big data}, \textit{data mining}). Many companies/people/organizations have their own internal definition for each of these terms e.g. one company's data scientist is another company's business analyst. The lack of a consistent term makes discussion challenging.

These terms are contentious partially because of the buzz associated with them and because of arguments about academic discipline subsetting e.g. some argue data science is a subset of statistics\footnote{``a data scientist is a statistician who lives in San Francisco" \citep{Priceonomics2017}} while others argue statistics is a subset of data science\footnote{\url{http://andrewgelman.com/2013/11/14/statistics-least-important-part-data-science/}}.  Academic funding (and egos) play a non-trivial role in the controversy. 

 Even the origin of the term is up for debate. We point to a few sources for developing\footnote{We don't claim this list is exhaustive.} both the literal string and the broader idea, ``data science" 
\begin{itemize}
\item \citep{tukey1962future}: the idea, not the literal string
\item \citep{naur1974computermethods}: the literal string, not the idea\footnote{Peter Naur uses the term ``data science" but in a narrower sense, focusing more on computation.}
\item \citep{chambers1993greater}: the idea, not the literal string
\item \citep{wu1998statistics}: both the idea and literal string
\item \citep{cleveland2001datascience}: both the idea and literal string
\end{itemize}

Others are credited with bringing the term/idea to industry e.g. \cite{patil2011building}. For more on the development of the term see \cite{donoho201750} and the two articles linked to below\footnote{\url{http://bulletin.imstat.org/2014/10/ims-presidential-address-let-us-own-data-science/} and \url{https://www.forbes.com/sites/gilpress/2013/05/28/a-very-short-history-of-data-science/}}. Both the idea (as defined in Section \ref{s:datasci}) and the literal string (A) have been around for a while and (B) have origins in statistics. 

The term data science has broken free from academic statistics into industry and other academic fields. In some cases, data science marginalizes the discipline of statistics which is a detriment to both statistics and anyone who analyzes data. Acknowledging the history of data science and statistics we hope will garner more respect for data science within statistics and for statistics from the broader community of data practitioners.

In the next few section we discuss the origins of data science, critiques of statistics and the broader notion of reproducibility.

\subsection{Critiques of statistics}
\label{ss:crit}

To first order, we summarize the critiques of statistics as: too much theory, not enough computation. We believe theory is important, however, too much theory at the expense of other things is a detriment to the field. See Section \ref{s:forward} for a more optimistic discussion of theory.

We thank Hongtu Zhu for pointing out two quotes from the Priceonomics website \citep{Priceonomics2017}\footnote{\url{https://priceonomics.com/whats-the-difference-between-data-science-and/}}, which provides some interesting discussion about data science and statistics. 

\begin{quotation}
Statistics was primarily developed to help people deal with pre-computer data problems like testing the impact of fertilizer in agriculture, or figuring out the accuracy of an estimate from a small sample. Data science emphasizes the data problems of the 21st Century, like accessing information from large databases, writing code to manipulate data, and visualizing data.
\end{quotation}

This quote is echoed by statisticians such as Hadley Wickham who lament the lack of priority academic statistics has given to such areas\footnote{E.g. see \url{https://priceonomics.com/hadley-wickham-the-man-who-revolutionized-r/} and the quote about ``The fact that data science exists as a field is a colossal failure of statistics."}. This point is well taken e.g. what proportion of statistics undergraduates are competent in R or Python? However, statistics is being sold short on its contributions to computation \citep{donoho201750}. For example, the R-Project \citep{Rproject2017} is a context where many members of the statistical community are directly engaged in writing and sharing code. Furthermore, there is a rather large area called \textit{statistical computing}, with quite a long history, see e.g. the American Statistical Association's Section\footnote{\url{http://stat-computing.org/computing/}}, which has been in continuous operation since 1972.

Another example is visualizing data. We acknowledge that a large fraction of the statistical community has created a culture of not devoting enough energy in the direction of looking at data. However, again there is a relatively small but very active community devoted to visualization, including the American Statistical Association's Section on Graphics\footnote{\url{http://stat-graphics.org/graphics/}}, formed in 1985.

Another popular, but incorrect belief is that statistics is not concerned with \textit{big data} (or phrased sans buzz words: statisticians do not care about computing things efficiently). As \cite{donoho201750} points out, statisticians have in fact always been interested in large data computation. For example, the word \textit{statistics} came about from work on census data which have been around for centuries and are large even by today's standards. The principle of \textit{sufficiency} is of course a mechanism to deal with large data sets efficiently. The point here is that these pursuits are a part of statistics, but are perhaps considered specialized as opposed to mainstream (e.g. in terms of publications in flagship journals, undergrad/graduate education, etc). 

A second quote from Priceonomics (also echoed by \cite{chambers1993greater, donoho201750})

\begin{quotation}
Statistics, on the other hand, has not changed significantly in response to new technology. The field continues to emphasize theory, and introductory statistics courses focus more on hypothesis testing than statistical computing... For the most part, statisticians chose not take on the data problems of the computer age.
\end{quotation}
This point is on target in a number of ways, but we take a different viewpoint on a few issues.

The first is on the value of statistical theory. Far from viewing it as something old fashioned and hence useless, we argue that the need for theoretical thinking is greater than ever in the data science age (see Section \ref{s:forward}). There are few calls yet for \emph{data science theory}, although the US National Science Foundation's TRIPODS program\footnote{\url{{https://www.nsf.gov/funding/pgmsumm.jsp/pimsid=505347}}} is an important exception. However we predict that as the field evolves there will be growing realization of the importance of that line of thought, as more and more algorithms become available with very little meaningful basis for making the critical choice among them.

The second is the statement that ``statistics courses focus more on hypothesis testing''. This makes the statistical outsider's mistake of thinking that statistics is a set of recipes for doing data analysis. It misses the deeper truth understood by people who practice data analysis: when properly taught, statistics courses teach an important way of thinking called the \emph{scientific method}. The main idea is that to be really sure of making actual discoveries (as opposed to finding spurious and ungeneralizable sampling artifacts) scientists should first formulate a hypothesis, then collect data and finally analyze.

One can be forgiven, however, for mistaking statistics as a set of recipes. Too many people interact with statistics exclusively via a standard Statistics 101 type class which may in fact treat statistics as a handful of formulas to memorize and steps to follow. While we believe the material taught in these courses is vital to doing science, it is perhaps time to rethink such introductory classes and teach data before (or concurrently with) teaching statistics. See Section \ref{ss:edu} and the references therein.

\subsection{Reproducibility and communication}
\label{ss:repcom}

The idea of \textit{reproducibility} has become an important topic in science \citep{peng2011reproducible, stodden2012reproducible, kiar2017comprehensive}. A narrow sense of reproducibility is \textit{the ability to recompute results} i.e. that someone else can easily obtain the data and run the code used in the original experiment \citep{leek2015opinion, patil2016statistical}. Reproducibility, in the broader sense, is about three things: scientific validity, communication and methodological development.

A number of factors go into assessing the scientific validity of a study. The gold standard of scientific validity is \textit{independent replication}. Replicability is about the ability for someone else to rerun the same experiment and obtain the same results\footnote{The literature is not consistent about the definitions of reproducibility and replicability. In this paper we use the definitions given here.}.  Reproducibility is about checking the details of the scientific argument made in a paper (much in the way a mathematician would check the details of a proof of a theorem). The requirement that the code which produced the results can be rerun at a later time and will create the exact same figures/numbers is an important condition for assessing the correctness of the results. It is also a software engineering issue; the scientist has to write code which is understandable, well documented, publicly available, and persists across time and computers\footnote{Writing code that continues to work overtime is non-trivial; it involves maintaining the same computing environment and managing dependencies correctly e.g. the software packages the code uses change over time, version 1.1.1 might work the same as version 2.1.1.}.

The analysis code is a key part of communicating the analysis itself, citing Jon Claerbout \citep{buckheit1995wavelab}

\begin{quotation}
[a]n article about computational science in a scientific publication is not the scholarship itself, it is merely advertising of the scholarship. The actual scholarship is the complete software development environment and the complete set of instructions which generated the figures.
\end{quotation}
The text summary of an analysis in the paper may not include all the details of a complex analysis (e.g. how were hyper-parameters tuned, how were the data preprocessed, etc). Moreover the description of the analysis may be incorrect: numbers can be misreported by accident or intentionally, the code may not work the way the analyst thought it did\footnote{Understanding the nitty-gritty details of how statistical software works is not trivial: how does the optimization routine determine convergence? Are the data mean centered by default? There is a lack in uniformity in how statistical software is written; we believe this is exacerbated by the lack of of statisticians writing statistical software.}, there may be bugs in the code, etc. 

Computational methodologies are often reused and built upon. If the code for a methodology is not available then the next person who wants to use/iterate on that methodology has to re-write the code\footnote{Even if the code for a study is available, someone may still want to rewrite the code say in another language. In this case have the original code available to base the new code on is helpful.} which is both an inefficient use of time/resources and can lead to errors.

Three of the main barriers to reproducibility are: culture, computational tools and education \citep{peng2011reproducible, stodden2012reproducible, leek2015opinion}. Even if the cultural and educational problems are solved, there are still technical challenges which discourage reproducibility in practice. These issues are primarily about software engineering; while reproducibility is technically achievable, it is often too burdensome in practice\footnote{Publishing messy code is still beneficial and certainly better than not publishing code \citep{barnes2010publish}}. For example, ``in a recent survey of the machine learning community--the single biggest barrier to sharing code and data was the time it takes to clean up and document the work to prepare it for release and reuse" \citep{stodden2012reproducible}. Other issues include: data sharing, data privacy, software sharing, software verification (e.g. unit testing), writing readable code, version control, and legality. 

Section \ref{s:compcom} discusses how the rise of \textit{literate programming} in data science has improved our ability to do and communicate reproducible science. It also discusses how the sharing of code and educational resources is a boon to the field.

\section{Some principal components of data science}
\label{s:modes}

In this section we discuss different approaches to data analysis and how they relate to broader trends in data science. We call these \textit{modes of variation} in the sense that they explain variation in the practice of data analysis. Each of the modes discussed below (e.g. predictive vs. inferential analysis) represents a spectrum between two methodologies, one more associated with data science and the other with classical statistics. These modes help explain why different communities seem to talk past each other and why some techniques (e.g. computation) have become more popular in recent times.

\subsection{Prediction vs. inference (do vs. understand)} \label{ss:predinfer}

A computer scientist might pejoratively describe a linear or logistic regression as shallow and quaint\footnote{The use of shallow means we can view a generalized linear model as a neural network with 0 layers. The more layers a network has, the more complex of a pattern it can find \citep{goodfellow2016deep}.}. A statistician might express bewilderment at the buzz around deep learning and question why more principled, interpretable statistical models won't do the trick. The point here is that these two imaginary, curmudgeonly academics are thinking about problems with different goals. The computer scientist is trying to build a system to accomplish some task; the statistician is typically trying to learn something about how the world works.

Prediction vs. inference is a spectrum. Many complicated problems have well defined subproblems which are closer to one end or the other end. The distinction we are trying to make here is maybe better described as engineering vs. science (or at least broad generalizations thereof). Engineering is the business of creating a thing that \textit{does} something\footnote{I.e. the output of a predictive model may be interesting insofar as it helps us do something.}. Science is the business of \textit{understanding} how something works. Of course engineers use science and scientists use engineering. But if we focus on the end goal of a particular problem we can probably, reasonably classify that problem as either science or engineering\footnote{In other words, in many cases understanding is primarily a means to and ends for predictive problems and visa versa.}.

\cite{breiman2001statistical} discusses many of the differences between predictive and inferential modeling. Predictive modeling often uses more sophisticated, computationally intensive models. This often comes with a loss in interpretability and general understanding about how the model works and what the data look like \citep{freitas2014comprehensible}. Predictive modeling also places less emphasis on theory/assumptions because there are fairly good, external metrics to tell the analyst how well they are doing (e.g. test set error). 

Predictive modeling is one of the main drivers of \textit{artificial intelligence} (AI). The fact that data can be used to help computers automate things is perhaps one of the most impactful innovations of recent decades. Early attempts at AI type applications involved primarily \textit{rules based systems} which did not use a lot of data \citep{russell2009modern}; modern AI systems are typically based on \textit{deep learning} and are extremely data hungry \citep{goodfellow2016deep}. Section \ref{sss:automation} discusses some of the ways statisticians can make important contributions to AI/predictive modeling. Section \ref{sss:mlprocess} discuss some of the ways statisticians/scientists may benefit from and reasons to be aware of machine learning.

\subsection{Empirically vs. theoretically driven}
\label{ss:empiricaltheory}

\begin{quotation}
Data science is exploratory data analysis gone mad. -- Neil Lawrence\footnote{From Talking Machines season 3, episode 5 \url{https://www.thetalkingmachines.com/}.}
\end{quotation}

Most quantitative fields of study do both theoretical and empirical work\footnote{This statement probably applies to non-quantitative fields. For example, some academics in comparative literature are more ``empirical" in the sense they examine a particular body of work, draw conclusions and possibly generalize/relate their conclusions to other bodies of work. Other people in comparative literature apply ``theoretical methods."} e.g. theoretical vs. experimental physics. Within statistics, we might contrast \textit{exploratory data analysis} (EDA) vs. \textit{confirmatory analysis} i.e. searching for hypotheses vs. attempting to confirm a hypothesis (in this section \textit{theory} refers to the scientific theory being tested). 

It used to be that statistics and science were primarily theory driven. A scientist has a model of the world; they design and conduct an experiment to assess this model; then use hypothesis testing to confirm the results of the experiment. With changes in data availability and computing, the value of exploratory analysis, data mining, and using data to generate hypotheses has increased dramatically \citep{fayyad1996advances, blei2017science}. EDA often prioritizes the ability to rapidly experiment which means computation can dominate the analysis.

EDA can become problematic when the analyst puts too much faith in its results i.e. when the analyst mistakes hypothesis generation for hypothesis confirmation. Every statistician can list problems with simply \textit{correlation mining} a data set (e.g. false discovery, sampling bias, etc). These problems don't mean the results are wrong, but rather they mean exploratory analysis provides much weaker evidence for a hypothesis than confirmatory analysis. The amount this matters is context dependent.

One problematic idea is that EDA can solve every problem. For example, in the controversially titled article, ``The End of Theory: The Data Deluge Makes the Scientific Method Obsolete" it was argued that EDA will replace the scientific method \citep{anderson2008end}. We disagree. This article is an extreme example of the broader attitude that correlation, and fancy models applied to large data sets, can replace causal inference and the careful, time intensive scientific method. The point that EDA can contribute to scientific applications is well taken, however, and this in fact is becoming more common \citep{blei2017science}. These applications likely raise many interesting and impactful research questions arising from the increased value in EDA (in all of GDS 1-6). 

\subsection{Problem first vs. hammer looking for a nail}
\label{ss:hammer}

Some researchers take a \textit{hammer looking for a nail} approach; the researcher has developed/studied a statistical procedure and then looks for problems where it might be applicable. Other researchers aim to solve some particular problem from a domain. Note that the former approach is strongly correlated with, but not equivalent to theoretical research (same for the latter and applied research). 

Both research approaches are valid and productive, however the balance in academic statistics may have shifted too far to the former (hammer) approach. For example, Rafael Irizarry has some interesting further commentary\footnote{\url{https://simplystatistics.org/2014/07/25/academic-statisticians-there-is-no-shame-in-developing-statistical-solutions-that-solve-just-one-problem/}} which is echoed by a number of others \citep{tukey1962future, breiman2001statistical, wasserman2014rise, donoho201750}. We include this section because data science is focused on problem solving and it is this problem solving which makes data analysis useful to other disciplines.

\subsection{The 80/20 rule}

The 80/20 rule of data analysis is:\footnote{\url{https://simplystatistics.org/2014/03/20/the-8020-rule-of-statistical-methods-development/}} 
\begin{quotation}
One of the under appreciated rules in statistical methods development is what I call the 80/20 rule (maybe could even by the 90/10 rule). The basic idea is that the first reasonable thing you can do to a set of data often is 80\% of the way to the optimal solution. Everything after that is working on getting the last 20\%.
\end{quotation}

Applying basic models to a data set often provides the most value (and/or solves the problem of interest much of the time). The 80/20 rule explains part of why a number of techniques have become more valuable than in the past and why the six areas of GDS emphasize previously undervalued areas. These include: data visualization, exploratory data analysis, data mining, programming, data storage/processing, computation with large datasets and communication. 

\section{Going forward}
\label{s:forward}

In this section we discuss a number of areas we believe are particularly promising for statistics research, communication and education. A pervasive theme in this section is the role of computation, defined broadly as in \cite{nolan2010computing}.

\subsection{Research}
\label{ss:research}

As data analysis becomes more valuable, existing statistical theory and methodology also become more valuable. Criticisms of statistical theory (e.g. Section \ref{ss:crit} and \cite{donoho201750}) are largely about ignoring other, less mathematically glamorous areas of statistics. 

In this section we highlight a number of (potentially) promising areas of statistics research. There are many ways in which one can do \textit{good} statistical research, including all of those discussed in \cite{tao2007good}. The list below is biased in favor of our tastes and areas which involve both theory and aspects of greater data science (from Section \ref{s:datasci}). Rather than diminishing the role of mathematical statistics, we emphasize that many of these areas will require novel contributions from mathematical statistics which itself should be broadened.

\subsubsection{Complex data and representation}
\label{sss:complex}

The recently fashionable area of Big Data has received a large amount of well deserved attention and has led to many serious analytic challenges. Currently less well understood is that a possibly bigger challenge is non-standard or \textit{complex data}. In particular, many modern data analytic scenarios involve non-standard data such as: networks, text, sound, images, movies, shapes, very high dimensional data, data living on a manifold, etc. In response to this challenge, \citep{wang2007object, Marron2014OODA} have proposed \textit{Object Oriented Data Analysis} (OODA).

A fundamental concept of OODA is that in complex scientific contexts, it is often not even clear what should be the atoms of the statistical part of the analysis, i.e. the experimental units. OODA provides terminology for  pivotal team discussions on this topic between domain scientists and statisticians which lead to an effective final analysis, via resolving \textit{what should be the data objects?} An interesting example of this in an image analysis context can be found in \cite{lu2014object}.

The OODA discussion goes beyond choice of experimental units also to include the key technical issue of \textit{data representation}. This includes both fairly standard statistical techniques such as data transformation, but also mathematically deeper issues such as the appropriate data space, as discussed for example in \cite{pizer2017object}. Much of the success of deep learning may come from its ability to automatically discover useful data representation \citep{bengio2013representation}; connections between OODA and \textit{representation learning} are unexplored and potentially fruitful directions for future research.

Complex data and OODA present many research opportunities such as: invention of powerful new tools for data practitioners, computational challenges, methodological developments, and developments in statistical theory. OODA often involves bringing in a number of mathematical disciplines such as differential geometry, topology, optimization, probability, etc, providing a sense in which data science should become a truly interdisciplinary endeavor. It is also an opportunity to greatly extend usage of the term \textit{mathematical statistics} to include many more mathematical areas beyond just the conventional probability theory.

\subsubsection{Robustness to unknown heterogeneity}

An even less widely acknowledged and studied challenge in data science is \textit{data heterogeneity}. This topic is very current to many modern sciences, where there has been a growing realization that complex data collection by a single lab tends to result in sample sizes that are inadequate to address many modern scientific issues of interest. 

The poster child for this problem may be cancer research, where the very diversity of the disease requires very large sample sizes in a context where the needed measurements are very labor, time and cost intensive. Such challenges have led to the formation of large multi-lab research consortiums. In the cancer world, a well known effort of this type is The Cancer Genome Atlas (TCGA) \citep{cancer2012comprehensive}. While great care is taken in such efforts to try to standardize laboratory protocols and many other aspects of the data collection, it is well known among all data collectors that there are always impossible to control biases that creep into the data set when data from different labs are combined. 

Dealing with this issue is the main challenge of data heterogeneity. It is clear that the standard statistical Gaussian model for noisy data is no longer appropriate. When data from different sources are combined, a much more appropriate statistical model is a Gaussian mixture, but this presents a major challenge to classical statistical approaches such as the likelihood principle.

Much more research is strongly needed on all aspects of data analysis for heterogeneous data, including methods, computation and theory. It is also clear that real progress is going to be made by approaches which truly integrate all three of those. More discussion of important early work in that direction can be found
in \citep{buhlmann2016magging, buhlmann2016mathematics, marron2017big}.

\subsubsection{Scalable and robust models}
\label{sss:scalerob}

Robustness issues are important for modern, large data applications (see \cite{huber2011robust, hampel2011robust, staudte2011robust, maronna2006robust} for an overview of this area). For the gains to be realized from robust statistical procedures for large data sets, these procedures need to be computationally efficient (or \textit{scalable}). We want models which are both scalable and robust. We point to \cite{bottou2016optimization} for an overview of important computational methods in machine learning.

Robustness presents a challenge for developing scalable models for large data because often robust methods are harder to compute\footnote{E.g. L2 regularized (ridge) linear regression  has a closed form while L1 regularization (LASSO) does not.}. Models which are both scalable and robust present an opportunity for research developments drawing on all of statistics, optimization and computer science. The work of \cite{aravkin2016smart} is a recent example in this vein. We point this area out because it requires joint reasoning based on both computation and statistical theory. 

\subsubsection{Automation and interpretability}
\label{sss:automation}

One of the biggest ways data is impacting society is by powering automation through machine learning. While data driven automation has a lot of potential to do good, recent years have brought new attention to its negative consequences \citep{zarsky2016trouble, o2017weapons, doshi2017towards, crawford2017trouble}. For example, \cite{bolukbasi2016man} demonstrate that natural language processing algorithms trained on google news text data learned offensive gender stereotypes. In a book titled ``Weapons of Math Destruction" data scientist Cathy O'Neil provides many examples of how automation can perpetuate and even reinforce existing societal inequalities \citep{o2017weapons}. 

Statistics is the discipline historically most concerned with the myriad of ways in which data can be misleading. The rise of automation presents new opportunities to the discipline. First, the decades of statistics' hard won knowledge about dealing with data is salient to many applications of data driven automation. Second, automation presents new technical challenges to statistics since machine learning often involves applying sophisticated modeling techniques to large, complex datasets. One of the primary technical challenges is interpretability.

\textit{Interpretability}, as discussed in \citep{lipton2016mythos, doshi2017towards}, is desirable for a number of reasons including: trust, causality, transferability, informativeness, fair and ethical decision-making, legality, and safety. Additionally, interpretability can mean different things e.g. a visualization, a verbal explanation, understanding the model, etc. The canonical examples of interpretable models are generalized linear models and decision trees. The canonical example of an \textit{uninterpretable} model is a deep neural network. One interesting approach to interpretability is discussed in \cite{ribeiro2016should}, called \textit{local interpretable, model-agnostic explanations} (LIME), which proposes the use of simple models to help explain predictions made by more sophisticated models. 

The pursuit of understanding machine learning models to make them more interpretable is a wide open area for statisticians. These questions may involve narrowing the gap between predictive and inferential modeling discussed in Section \ref{ss:predinfer}.

\subsubsection{Machine learning and data processing}
\label{sss:mlprocess}

Complex data preprocessing pipelines are familiar to statisticians working in many areas including genomics and neuroscience \citep{gentleman2006bioinformatics, kiar2017comprehensive}. The raw data go through a number of processing steps before reaching the analyst as a .csv file. These steps may include: simple transformations, complex algorithms, output of statistical models, etc. Knowledge of the processing pipelines can be important for the analyst because the pipelines can have errors and introduce systemic biases in the data. 

Machine learning, particularly deep learning, may cause complex data processing pipelines to be more frequently used in modern data analysis. A lot of work in machine learning is about measurement: image recognition, image captioning, speech recognition, machine translation, syntactic parsing, sentiment analysis, etc \citep{goodfellow2016deep}. These examples some what blur the line between data processing and data gathering. We suspect that it will be more common for data analysts to have one or more variables in their datasets which are the output of a deep learning model. While machine learning can provide the analyst with rich, new variables, it should also give the statistician some pause. 

There are at least two major issues with using a deep learning model to gather data: it puts black box models into the data processing pipeline and introduces an external data set (i.e. the one used to train the deep learning model). Neither of these are new issues, but they will become more prominent. The black box model is problematic because it means the analyst cannot fully understand where the raw data came from. External data is problematic because it can infect new datasets it comes into contact with via the deep learning model on which it was trained\footnote{For example, if google's algorithms mistakenly think someone is dead, then likely the rest of the world will too \url{https://www.nytimes.com/2017/12/16/business/google-thinks-im-dead.html}}. Both of these issues are likely to create systemic biases which are challenging to detect and appropriately handle.

We suspect that machine learning will create a lot of value in data processing in many domains. However there are important theoretical and methodological questions which will arise when using deep learning models to gather data for inferential, scientific applications.  

\subsection{Computation and communication}
\label{s:compcom}

In this section we discuss a number of ways in which computation can improve communication in data analysis\footnote{Our argument is that computation can help communication. Others have taken this idea further and use computation, specifically information theory, as a metaphor for communication e.g. \citep{doumont2009trees}.}. Section \ref{sss:LP} below discusses how \textit{literate programming} helps solve many of the issues in reproducibility discussed in Section \ref{ss:repcom} above.

\subsubsection{Literate programming}
\label{sss:LP}

\textit{Literate programming} is a concept developed by computer scientist Donald Knuth in the 1980s \citep{knuth1984literate}.

\begin{quotation}
Literate programming is a methodology that combines a programming language with a documentation language, thereby making programs more robust, more portable, more easily maintained, and arguably more fun to write than programs that are written only in a high-level language. The main idea is to treat a program as a piece of literature, \textbf{addressed to human beings rather than to a computer}. 
\end{quotation}

Literate programming is about both cultural and technical developments. Cultural in the sense that it emphasizes the programmer's responsibility to communicate with humans. Lessons learned in English courses about organization, clarity, prose, etc are relevant in computer science. Technical in the sense that programming languages need to be altered or created to enable the programmer to write documents which communicate effectively \citep{knuth1984literate}. Literate programming is important to data analysis because data analysis is done with computer code; explaining the results and details of an analysis involves communicating the details of a computer program. 

Traditional data analysis code is not written for human consumption (e.g. little documentation, not made public); the analyst communicates the results through prose and figures which summarize the execution of the code. For some applications this is acceptable in the sense that it accomplishes what is needed. Problems with this style of programming include: it's not reproducible, it's not generalizable, it doesn't communicate the details of the analysis, it can make it harder to find errors with the analysis code, etc. As data analysis becomes more used, more multidisciplinary, and more complex these issues only become more important. The core idea of literate programming -- writing code whose target audience includes humans -- has a lot to contribute to data analysis. 

Communicating data analysis through the medium of code is challenging. Best practices have not yet been established (e.g. how should the code be commented, how to evaluate trade-offs between code clarity and simplicity/efficiency). Software engineers have built up best practices for programming \citep{wilson2014best, wilson2017good}, which can be helpful for data analysts to learn, but may need to be adapted in some cases\footnote{For example, it is often suggested that code comments should describe \textit{why} the code was written the way it was, not \textit{what} the code is doing. For data analysis, where the target audience is probably less experienced with programming, describing the \textit{what} may also be useful.}. In addition to methodology, there are technical developments which make literate programming easier.

We highlight \textit{knitr} \citep{xie2015dynamic} and R Markdown\footnote{For more information and examples see \url{http://rmarkdown.rstudio.com/}.} as examples of research into GDS 6, ``science about data science." As discussed in \cite{donoho201750},

\begin{quotation}
This helps data analysts authoring source documents that blend running R code together with text, and then compiling those documents by running the R code, extracting results from the live computation and inserting them in a high-quality PDF file, HTML web page, or other output product. In effect, the entire workflow of a data analysis is intertwined with the interpretation of the results, saving a huge amount of error-prone manual cut and paste moving computational outputs and their place in the document. 
\end{quotation}

There are other examples of technological developments in literate programming for data analysis (e.g. Jupyter notebooks \cite{perez2015project}). Others\footnote{E.g. see the list of people discussed in \url{https://simplystatistics.org/2015/12/11/instead-of-research-on-reproducibility-just-do-reproducible-research/}} developing tools to solve technical issues\footnote{The complexity and time costs to making research reproducible is, in part, technical issue.}  in reproducible research include \citep{sandve2013ten, kiar2017comprehensive}. Literate programming is also impactful in statistics education\footnote{\url{https://simplystatistics.org/2017/06/13/the-future-of-education-is-plain-text/}}, particularly for demonstrating programming examples\footnote{E.g. see each of the notes from \url{https://idc9.github.io/stor390/}.}. 


\subsubsection{Open source}
\label{ss:opensource}

As discussed in Section \ref{ss:repcom} reproducibility has a number of technical challenges and literate programming goes a long way to address these issues. Reproducibility crucially demands that the analysis code and data be publically released. This is probably more of a cultural issue than a technical issue (though it's worth pointing out that technologies such as GitHub make sharing code much easier). Some journals (e.g. Biostatistics) encourage the authors to release their code; we hope that more journals will follow this example.

Releasing analysis code makes the original analysis more impactful. A statistics paper is useful primarily to statisticians; an R package is useful to anyone who knows R. If a future analyst has existing code to work with then their life will often become easier. This can take a number of different forms depending on the project. 

For some projects, writing an R script which carries out the analysis will allow a future analyst to copy, paste and modify the original script. In other cases, if a researcher develops a new algorithm or complex methodology then releasing a software package may be most appropriate. Resources such as \citep{wickham2015r} make developing software packages fairly straightforward. Finally, many research advancements build upon existing algorithms/methodologies. In this case, enhancing an existing software package (as opposed to developing a new one) can be most cost effective. The open source software community has built up best practices for developing and maintaining open source software projects\footnote{E.g. see \url{https://github.com/scipy/scipy/blob/master/HACKING.rst.txt}.}. 

It's worth noting there is a large amount of open source statistical software already available, e.g. CRAN, bioconductor, scipy, sklearn. However, we suspect the majority of modern statistics research does not make it into good, publicly available software packages.

Open source extends beyond analysis/research code to educational resources. There are now many technologies which make it easy to publish educational resources for free through a variety of mediums: books, blogs, massive open online courses (MOOCs\footnote{\url{http://mooc.org/}}), websites, etc. As the first author can attest, data science has been particularly good at making resources available for free online. This is an important trend for a number of reasons; it saves students money, provides them with more resources to learn from and it democratizes the subject in the sense that anyone can learn the basic skills to get started in the subject. 

The technology is largely available to make most data analyses, statistical algorithms and educational resources freely accessible. Two of the biggest barriers to releasing these materials are education and incentives. Education in the sense that many data analysts don't currently know these technologies which causes the release of code to be too burdensome. Incentives in the sense that there are few of them to encourage releasing these materials. How would a statistics department's hiring committee value a software package vs. a paper? Surely they are both intellectually challenging to produce and valuable, but we suspect the latter would typically receive much more weight that the former. There are some examples of people promoting the value of research software \citep{sonnenburg2007need} and we hope there will be more in the future. 

While open source has lot of potential value, it also raises some questions. For example, not all data can be shared due to privacy/confidentiality reasons; what steps should researchers take in this case to share their analysis code/data? One research avenue worth noting in this regard is \textit{differential privacy} \citep{smith2016differentially}. As noted in, \citep{graves2000predicting, eick2001does} software decays if it is not actively maintained. This raises the question, for how long should research be reproducible\footnote{\url{https://simplystatistics.org/2017/02/01/reproducible-research-limits/}}?

\subsection{Education}
\label{ss:edu}
A number of people have written about updating the statistics curriculum in ways which better reflect the broader definition of data science and the skills required for doing data analysis \citep{nolan2010computing, american2014curriculum, de2017curriculum, hardin2015data, hicks2017guide}. A number of programs which have embraced these recommendations have proven to be successful such as the Johns Hopkins Data Science Specialization on Coursera\footnote{\url{https://www.coursera.org/specializations/jhu-data-science}} \citep{kross2017democratization} and Berkeley's Data8 program\footnote{\url{http://data8.org/}} \citep{alivisatos2017stem}. We observe three takeaways from this literature (and our own experiences): more computation, more data analysis and the use of open source material. 

Communication is another area worth highlighting in this education section because of its importance and ubiquity. The modern data analyst is expected to communicate across a number of different media: written papers/reports, static/dynamic visualizations, online via creating a website/blog, through code, etc. Communication is already under represented in STEM education, in spite of the demand from employers (academic and industrial alike) \citep{felder2016teaching}. Both authors have made effort\footnote{E.g. including a lecture on communication in an undergraduate data science course: \url{https://idc9.github.io/stor390/notes/communication/communication.html}} to include communication in statistics education \citep{marron1999effective}.

\subsubsection{More computation}

Computation comes into data analysis in a number of ways, from processing data to fitting statistical models to communicating the results. For example, as discussed in Section \ref{ss:repcom} many of the issues in reproducibility can be partially addressed by teaching more software engineering to scientists. Many new areas of statistics research involve tackling both statistical and computational issues e.g. see Section \ref{sss:scalerob} on scalable, robust estimators or \citep{efron2016computer}.

With the large number of technologies involved with data analysis (programming languages, visualization software, algorithms, etc) one often feels a bit overwhelmed at what one might be expected to know. It is infeasible to know everything. This is where updating the statistics education curriculum is critical. There is probably some rough core set of computational knowledge every trained statistician should have. Once the fixed cost of learning the core computational curriculum is paid, the marginal cost of learning additional computational skills will go down.

\subsubsection{Pedagogy}

The references cited above in Section \ref{ss:edu} make a number of good recommendations for improving statistics education. We highlight a couple of points here. 

As many of the above references discuss, the current statistics curriculum often lacks data analysis \citep{tukey1962future, nolan2010computing}. Real data analysis makes the discipline more concrete to students. Focus on solving a real problem can be engaging to students who might otherwise find the subject boring. Teaching data analysis is challenging\footnote{\url{https://simplystatistics.org/2017/12/20/thoughts-on-david-donoho-s-fifty-years-of-data-science/}}, but it's challenging in the way that teaching the practice of engineering or using the scientific method is challenging. By not giving students practice doing data analysis for a real problem, the statistics curriculum may encourage students to view statistical methodology as a hammer to be procedurally applied to data. It's well established in engineering and the physical sciences that students should get some practical experience doing the thing during their education: why does the same principle not apply more often statistics at the undergraduate (and graduate) level?

When teaching statistical modeling\footnote{E.g. in an upper level undergraduate course such as UNC's STOR 455: Statistical Methods I.} it might be more effective to first introduce the model (e.g. linear/logistic regression) in terms of a predictive context instead of the traditional inferential context. While the math behind linear models may not be particularly sophisticated, the concept of randomness and relating it to the real world through data is non-trivial. Moreover, inferential modeling comes with a lot of baggage. For predictive purposes, models such as linear regression can be introduced as an optimization problem which can be heuristically motivated and analytically solved\footnote{See for example \url{https://www.inferentialthinking.com/chapters/13/prediction.html}.}. Once students are comfortable with data modeling we can then introduce inferential/confirmatory modeling.

Finally, we suggest teaching data before statistics in introductory statistics courses. In other words, we should teach exploratory analysis before inferential analysis. This would involve teaching programming, data visualization and manipulation before teaching hypothesis testing. Students are more likely to care about hypothesis testing if they have actually worked on a real problem/dataset which motivates it (as opposed to a hypothetical or artificial one). Berkeley's new course Data 8: The Foundations of Data Science appears to take this approach\footnote{E.g. see the order of the chapters in the textbook: \url{https://www.inferentialthinking.com/}.}.

\section{Conclusion}

So far the arguments in this paper have been about providing value to society by broadening the discipline in technical ways. Equally as important is increasing diversity in statistics by encouraging women and underrepresented minorities to join and stay in the discipline. Programs, conferences and other efforts such as: ASA Committee on Minorities in Statistics\footnote{\url{http://community.amstat.org/cmis/home}}, the Women in Machine Learning conference\footnote{\url{http://wimlworkshop.org/}} and the Women in Statistics and Data Science conference\footnote{\url{http://ww2.amstat.org/meetings/wsds/2018/}} should be strongly encouraged and expanded upon.

We return to the question of whether data science and statistics are really two different disciplines. If statistics is defined as the narrow discipline described by the quote from John Chambers in Section \ref{s:datasci} then the answer is yes. However, if statistics embraces the broader idea of greater data science (e.g. by putting more focus on computation in education, research and communication) then we argue the answer is no.



\begin{acknowledgements}

This research was supported in part by the National Science Foundation under Grant No. 
1633074. We would like to thank Deborah Carmichael for editorial comments. 

\end{acknowledgements}

\bibliographystyle{spbasic}      

\bibliography{refs}   

\end{document}